\documentclass[11pt,twoside]{article}

\usepackage{asp2006}
\usepackage{epsf}
\usepackage{psfig}
\usepackage{lscape}

\begin{document}
\title{Monte Carlo Radiative Transfer Simulations: Applications to
Astrophysical Outflows and Explosions} 
\author{S. A. Sim$^{1}$, M. Kromer$^{1}$, F. K. R\"{o}pke$^{1}$,
  E. I. Sorokina$^{2}$, S. I. Blinnikov$^{1,2}$, D. Kasen$^{3}$, W. Hillebrandt$^{1}$}
\affil{$^{1}$Max-Planck-Institut f\"{u}r Astrophysik, Karl-Schwarzschild-Str. 1, 85741 Garching, Germany\\
$^{2}$Institute for Theoretical and Experimental Physics,
  Bolshaya Cheremushkinskaya 25, 117218 Moscow, Russia\\
$^{3}$University of California Santa Cruz, 1156 High Street, Santa Cruz, CA 95064, USA}   

\begin{abstract} 
The theory of radiative transfer provides the link between the
physical conditions in an astrophysical object and the observable
radiation which it emits. Thus accurately modelling radiative
transfer is often a necessary part of testing theoretical models by
comparison with observations.  We describe a new radiative transfer
code which employs Monte Carlo methods
for the numerical simulation of radiation transport in expanding
media. We discuss the application of this code to the calculation
of synthetic spectra and light curves for a Type Ia supernova
explosion model and describe the sensitivity of the results to certain
approximations made in the simulations. 
\end{abstract}

\section{Monte Carlo Radiative Transfer}

Since almost all we know about astronomical objects is inferred from
the radiation which they emit, the theory of radiation transport often
has a key role in testing our understanding of astrophysics. Although
there are a variety of competitive approaches used for radiative
transfer simulations, Monte Carlo methods are particularly well-suited
for many modern astrophysical applications. In the Monte Carlo
approach, the radiation field is discretized into quanta which
represent bundles of photons. By propagating these quanta through a
model of an astrophysical object, and simulating their interactions,
synthetic spectra and light curves can be obtained. 
This method has the particular advantage that matter-radiation
interactions are always treated locally meaning that
multi-dimensionality, time-dependence and large-scale velocity fields
can all be incorporated readily. 

\section{The ARTIS code}

Here we describe a new radiative transfer code 
({\sc artis}; Kromer \& Sim 2009) which has been developed for
application to Type Ia supernova (SN Ia) explosion models. The code is
based on a Monte Carlo {\it indivisible packet} scheme described by
\cite{lucy02,lucy03,lucy05} and was developed from the grey radiative
transfer code of \cite{sim07}. 

The code is designed to simulate time-dependent, three-dimensional 
radiation transport in supernova ejecta during
the phase of homologous expansion. 
The optical display of SNe~Ia is powered by the
energy released in the radioactive decay of isotopes synthesized
during the explosion, predominantly $^{56}$Ni and its daughter nucleus
$^{56}$Co. 
Therefore, the code starts from an 
initial distribution of $^{56}$Ni in the ejecta and then the
subsequent radioactive decays are followed. These decays initially
give rise to gamma-ray photons which, at least for early epochs
when the ejecta are optically thick, are rapidly down-scattered and
absorbed by photoelectric processes. 
This heats the ejecta.
The subsequent re-emission of
ultraviolet, optical and infrared emission by the ejecta is then
simulated to obtain spectra and light curves. 
The code does not assume local thermodynamic equilibrium (LTE) but
includes an approximate non-LTE (NLTE) treatment of ionization and a
detailed approach to line scattering and fluorescence.
For a complete description of the code see \cite{sim07} and \cite{kromer09}.

\section{Testing ARTIS with a standard explosion model}

To test the {\sc artis} code we have performed a variety of radiative
transfer simulations for the well-known W7 SN~Ia explosion model
\citep{nomoto84,thielemann86}. Although this one-dimensional model is 
considerably simpler than modern three-dimensional explosion models
(e.g. R\"{o}pke \& Niemeyer 2007; R\"{o}pke et al. 2007),
it is known to predict spectra and light
curves in reasonable agreement with observations
(e.g. Jeffery et al. 1992; H\"{o}flich 1995; Nugent et al. 1997; Lentz
et al. 2001; Baron et al. 2006; Kasen et al. 2006)
and therefore
provides a realistic test model for our radiative transfer simulations.

For our W7 test simulations, the explosion model properties 
(ejecta density, composition and initial distribution of radioactive
isotopes) were mapped to a homologously expanding 50$^3$ Cartesian
grid and the expansion followed for 100 logarithmically-spaced time
steps spanning the time interval from 2 to 80 days after
explosion. 
The assumption of homologous expansion in the W7 model has been tested
using the {\sc stella} code in a manner similar to that described by
\cite{woosley07}. That test showed that the density
structure is affected only slightly during the first weak after
explosions and that thereafter homologous expansion becomes a very
good approximation.
The propagation of a 
total of five million Monte Carlo energy packet quanta were simulated
from their initial release by the radioactive decay of either
$^{56}$Ni or $^{56}$Co until, after multiple radiation-matter
interactions, they escaped from the computational domain as bundles of
ultraviolet, optical or infrared (UVOIR) photons. The escaping
packets were then binned by time of escape and photon frequency to
construct time-dependent spectra for the model. Since the W7 model is
one-dimensional, it is unnecessary to bin the escaping quanta
based on direction of scape but this can be readily done to
obtain viewing-angle dependent spectra for multi-dimensional models.

\subsection{Synthetic spectra from ARTIS}

Figure~1 shows a sequence of three optical spectral snapshots obtained with
the {\sc artis} code for the W7 model. 
A convenient property of Monte Carlo radiative transfer simulations is
the ease with which the propagation histories of the Monte Carlo
quanta can be used to understand the manner in which the spectra
features are formed. For each escaping quantum, {\sc artis} records
the details of its last radiation-matter interaction with either a
bound-bound, bound-free or free-free event. This information can
then be used to identify the processes responsible for features in
the spectra. In Figure~1, the areas above and below the total spectrum
are shaded
to indicate the atomic number of the elements which last affected the
escaping quanta for each wavelength bin.
This makes it easy to understand
how the spectral features are formed and one can readily see how the
contributions to the spectra evolve in time. For example, in Figure~1 it is
clear from the shading that the early phase spectra are strongly
affected by intermediate-mass elements such as Si and S while the
later time spectra are dominated by elements of the iron group. This
is a well-known 
consequence of the layered structure of the W7 model. At early times,  
the
outer layers (which are rich in the products of partial nuclear
burning) are optically thick. As time passes, however, the expansion
of the ejecta causes these layers to become optically thin such that
radiation escapes directly from the inner region which is dominated by
iron group material.

\begin{figure}[h]
\plotfiddle{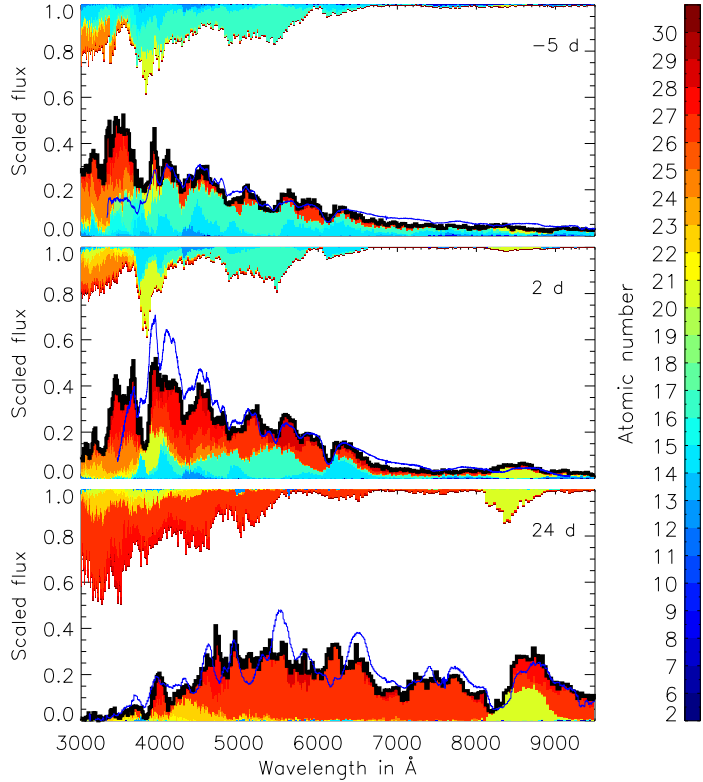}{9.3cm}{90}{85}{85}{500}{-100}
\caption{Computed spectra for the W7 model for three epochs: -5, +2
  and +24 days measured relative to the maximum of the B-band light
  curve (which occurs around 19 days after explosion). The emitted
  model spectra are shown as the heavy black histogram in each
  panel. The area under the curve is shaded to indicate the elements
  responsible for the last bound-bound interactions of escaping Monte
  Carlo quanta in the corresponding frequency bin.
  The shaded region at the top of each panel identifies
  which elements were last responsible for removing packets from the
  corresponding frequency bin (this can be used to identify absorption
  features in the spectra).
  The observed spectra of SN 1994D (Patat et al. 1996) 
  are shown for comparison (solid blue lines). These have not been
  corrected for redshift or extinction.}
\end{figure}

For comparison, the observed
spectra of a fairly normal SN~Ia (SN~1994D, Patat et al. 1996) are
over-plotted for the same epochs. 
The overall flux distribution and 
general properties of the observed spectral features (e.g. the
characteristic Si~{\sc ii} line at 6355\AA~ and the Ca~{\sc ii}
infrared triplet at 8549\AA) are reasonably
well-reproduced by the model, suggesting that the numerical
simulations capture much of the necessary physics required for the
interpretation of the observations. There are, however, some clear
discrepancies between the observational data and the synthetic spectra
(e.g. excess emission below $\sim$4000\AA~ in the early-time model spectra
and an extra emission feature around $\sim$6500\AA~ in the later time
spectrum) -- but some disagreement is expected since the W7 model has not been
fine-tuned to match observations of any specific SN
in detail.

\subsection{Sensitivity to plasma conditions}

Multi-dimensional radiative transfer
simulations for SNe~Ia require significant
computational resources and approximations must currently be made to
make the simulations feasible. Two particularly important issues are
the completeness of the atomic data set used
and the 
sensitivity of the synthetic observables to the treatment of the
plasma conditions (particularly the excitation/ionization state).
We have therefore used the W7 model as a standard case to investigate
some of these effects with {\sc artis}; photometric light curves computed from
several of our numerical simulations are shown in Figure~2. 
We also compared our results
with those of other SN radiative transfer codes to quantify the extent
to which different numerical methods affect the synthetic
observables.

\begin{figure}[h]
\plotfiddle{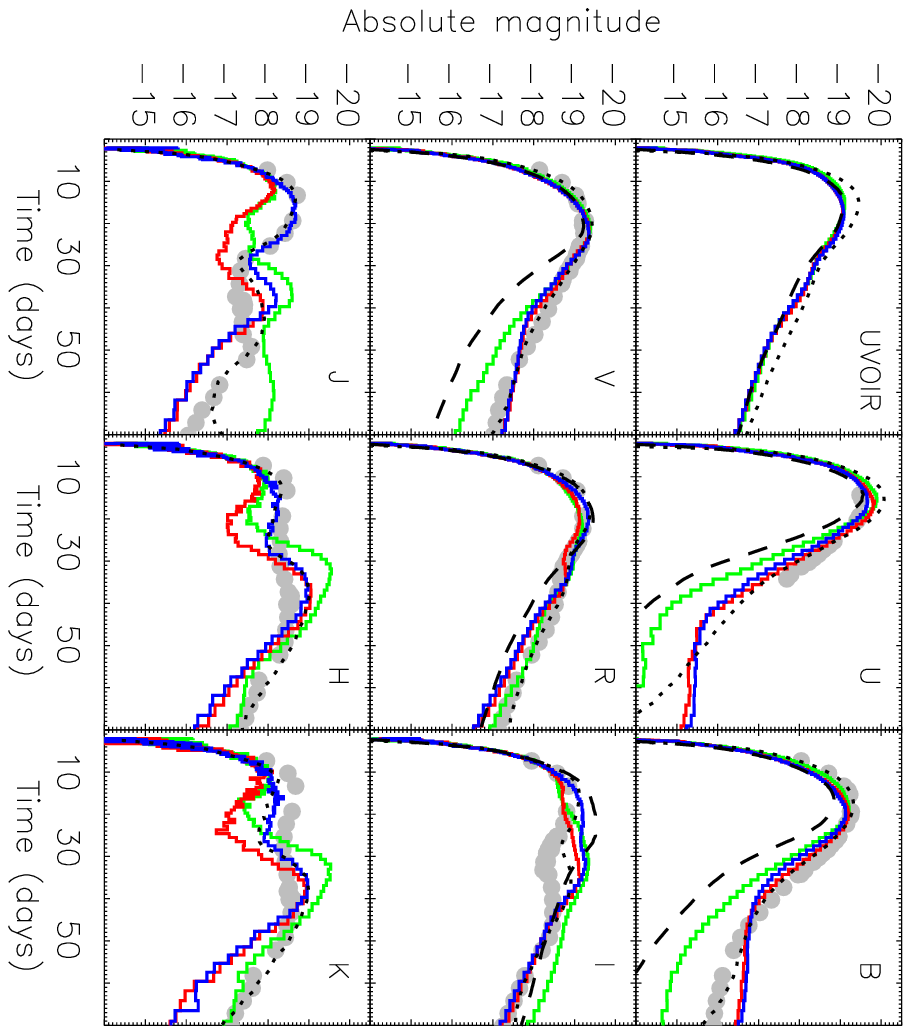}{7.2cm}{90}{88}{88}{495}{-100}
\caption{Computed light curves for the W7 model. The panels show the
  UVOIR bolometric, U, B, V, R, I, J, H and K band light curves
  obtained from our time-dependent spectra using the filter functions
  of Bessell (1990) and Bessell \& Brett (1988). The solid histograms
  show the light curves obtained with {\sc artis}
  using (i) our approximate NLTE treatment of ionization and an atomic
  data set containing $\sim 8 \times 10^6$ atomic lines (blue), (ii)
  NLTE with only $\sim
  4 \times 10^5$ lines (red) and (iii) an LTE treatment of ionization
  with  $\sim 4 \times 10^5$ lines (green). For comparison we show
  observations of SN~2001el (Krisciunas et al. 2003; grey circles) and
  light curves obtained with the {\sc sedona} (dotted lines) and
  the {\sc stella} codes (dashed lines; UVOIR, U, B, V, R, and I).} 
\end{figure}

Using the {\sc sedona} code, \cite{kasen06b} showed that realistic
atomic data sets containing many millions of spectral lines 
are required to accurately model radiation transport in SN~Ia. This is
particularly true at NIR (near-infrared) wavelengths where the spectrum can
be significantly affected by fluorescent emission in forests of weak
lines of iron-group elements. This is confirmed by our simulations with
{\sc artis}. Figure 2 shows light curves computed with our approximate
NLTE treatment of ionization but adopting different atomic data
sets drawn from the line lists computed by Kurucz \& Bell (1995) and
Kurucz (2006). When we use a reasonably large atomic line list ($\sim 8 \times
10^6$ lines) we obtain NIR light curves which are significantly
brighter around their first maximum than with an atomic data set
restricted to only $\sim 4 \times 10^5$ lines. The optical light
curves are much less affected although there is a tendency
for the U, B and V bands to be slightly fainter owing to the flux
redistributed from these bands to the NIR. We note that the light
curves computed with the larger atomic data set are also in
quantitatively better agreement with observations (see Figure 2).

In Figure 2 we also show light curves computed with a pure LTE
treatment of the excitation/ionization state of the ejecta. These
light curves were obtained using the smaller atomic data set
($\sim 4 \times 10^5$ lines) mentioned above. For early times (up to
around 30 days in the optical bands and 20 days in the NIR bands),
these light curves agree well with those obtained with our NLTE
implementation. This is expected since LTE should be a good
approximation when the radiation is strongly trapped. At later times,
however, departures from LTE become strong and our NLTE treatment of
ionization predicts significantly higher ionization states throughout
much of the ejecta. This directly affects the observables causing the
U, B and V band to remain significantly brighter than suggested by
LTE. This illustrates the sensitivity of the observations to the
ejecta properties and highlights the need for a realistic treatment of
the plasma conditions if detailed comparisons to observations are to be
made.

\subsection{Comparison of results from different codes}

The agreement between the {\sc artis} light curves and those obtained
by {\sc sedona} \citep{kasen06a} and {\sc stella}
\citep{blinnikov02,blinnikov06} is encouragingly good (Figure 2).
Compared to {\sc sedona}, the {\sc artis} light curves computed
with the larger atomic data set agree very well in all bands up to
several weeks after maximum light. The difference which manifest at
later times are most likely attributable to difference in the manner in which
the codes treat the plasma conditions (see Kromer \& Sim 2009 for
further discussion). The current version of {\sc stella} adopts an LTE
treatment of the plasma conditions with photon redistribution modelled
using an approximate source function. 
As expected, its light curves
are similar to those obtained with the LTE implementation in
{\sc artis}. The {\sc stella} light curves shown here were computed
with an extended atomic data set containing 2.6 $\times$ 10$^7$
lines. This atomic data set improves aspects of the 
comparison with {\sc stella} relative to that shown in figure~7 of Kromer \& Sim
(2009; the {\sc stella} curves there used only 1.6 $\times$ 10$^5$
lines). In particular, the initial rise of the {\sc stella} light curves is faster.

\section{Future prospects}

Our results obtained from the W7 model suggest that our radiative
transfer simulations are able to produce realistic synthetic spectra
and light curves as required for the testing of SNe~Ia explosion
models. We have already used the {\sc artis} code to investigate
simple aspherical toy models \citep{kromer09,kromer09a} and will in
the near future use it to compute synthetic observables for
state-of-the-art hydrodynamical explosions models in order that their
predictions can be directly tested against observational data.


\begin{thebibliography}{}

\bibitem[Baron et al.(2006)]{baron06}
Baron  E., Bongard  S., Branch  D., Hauschildt  P. H., 2006, ApJ, 645,
480
\bibitem[Bessell(1990)]{bessell90}
Bessell M. S., 1990, PASP, 102, 1181
\bibitem[Bessell \& Brett(1988)]{bessell88}
Bessell M. S., Brett J. M., 1988, PASP, 100, 1134
\bibitem[Blinnikov \& Sorokina(2002)]{blinnikov02}
Blinnikov  S., Sorokina  E., 2002, preprint (arXiv:astro-ph/0212567)
\bibitem[Blinnikov et al.(2006)]{blinnikov06}
Blinnikov  S. I., R\"{o}pke  F. K., Sorokina  E. I., Gieseler  M., Reinecke  M., Travaglio  C., Hillebrandt  W., Stritzinger  M., 2006, A\&A, 453, 229
\bibitem[H\"{o}flich(1995)]{hoeflich95}
H\"{o}flich  P., 1995, ApJ, 443
\bibitem[Jeffery et al.(1992)]{jeffery92}
Jeffery  D. J., Leibundgut  B., Kirshner  R. P., Benetti  S., Branch  D., Sonneborn  G., 1992, ApJ, 397, 304
\bibitem[Kasen(2006)]{kasen06b}
Kasen  D., 2006, ApJ, 649, 939
\bibitem[Kasen et al.(2006)Kasen, Thomas, \& Nugent]{kasen06a}
Kasen  D., Thomas  R. C., Nugent  P., 2006, ApJ, 651, 366 
\bibitem[Kromer \& Sim(2009)]{kromer09}
{Kromer}, M. \& {Sim}, S.~A., MNRAS, 2009, 398, 1809
\bibitem[Kromer et al.(2009)Kromer, Sim, \& Hillebrandt]{kromer09a}
{Kromer}, M., {Sim}, S.~A. \& Hillebrandt W., 2009, in AIP Conf.
Ser. Vol. 1111, Probing Stellar Populations out to the Distant
Universe: Cefalu 2008, ed. G.~Giobbi, A.~Tornambe, G.~Raimondo, M.~Limongi,
L.~A.~Antonelli, N.~Menci, \& E.~Brocato (New York: AIP), 277 
\bibitem[Krisciunas et al.(2003)]{krisciunas03}
Krisciunas  K.  et al., 2003, AJ, 125, 166
\bibitem[Kurucz \& Bell(1995)]{kurucz95}
Kurucz R., Bell B., 1995, Atomic Line Data, Kurucz CD-ROM
No. 23. Smithsonian Astrophysical Observatory, Cambridge, MA
\bibitem[Kurucz(2006)]{kurucz06}
 Kurucz R. L., 2006, in EAS Publ. Ser. Vol. 18,Radiative Transfer and
 Applications to Very Large Telescopes, ed. P. Stee (EDP
 Science: Les Ulis), p.129
\bibitem[Lentz et al.(2001)]{lentz01}
Lentz  E. J., Baron  E., Branch  D., Hauschildt  P. H., 2001, ApJ, 557, 266 \bibitem[Lucy(2002)]{lucy02}
Lucy L. B., 2002, A\&A, 384, 725
\bibitem[Lucy(2003)]{lucy03}
Lucy L. B., 2003, A\&A, 403, 261
\bibitem[Lucy(2005)]{lucy05}
Lucy L. B., 2005, A\&A, 429, 19
\bibitem[Nomoto et al.(1984)Nomoto, Thielemann, \& Yokoi]{nomoto84}
Nomoto  K., Thielemann  F.-K., Yokoi  K., 1984, ApJ, 286, 644
\bibitem[Nugent et al.(1997)]{nugent97}
Nugent  P., Baron  E., Branch  D., Fisher  A., Hauschildt  P. H.,
1997, ApJ, 485, 812
\bibitem[Patat et al.(1996)]{patat96}
Patat  F., Benetti  S., Cappellaro  E., Danziger  I. J., della Valle
M., Mazzali  P. A., Turatto  M., 1996, MNRAS, 278, 111
\bibitem[R\"{o}pke \& Niemeyer(2007)]{roepke07a}
R\"{o}pke F. K., Niemeyer J. C., 2007, A\&A, 464, 683
\bibitem[R\"{o}pke et al.(2007)R\"{o}pke, Woosley, \& Hillebrandt]{roepke07b}
 R\"{o}pke F. K., Woosley S. E., Hillebrandt W., 2007, ApJ, 660, 1344
I. J., Patat  F., Turatto  M., 2001, MNRAS, 321, 254
\bibitem[Sim(2007)]{sim07}
Sim  S. A., 2007, MNRAS, 375, 154 
\bibitem[Thielemann et al.(1986)Thielemann, Nomoto, \& Yokoi]{thielemann86}
Thielemann  F.-K., Nomoto  K., Yokoi  K., 1986, A\&A, 158, 17
\bibitem[Woosley et al.(2007)]{woosley07}
Woosley S. E., Kasen D., Blinnikov S., Sorokina E., 2007, ApJ, 662, 487
\end{thebibliography}
\end{document}